\documentclass[11pt]{article}

\usepackage[margin=2.5cm]{geometry}

\usepackage[english]{babel}
\usepackage[T1]{fontenc}
\usepackage[]{times}
\usepackage{comment}

\usepackage{amsmath}
\usepackage{amssymb}
\usepackage{graphicx}
\usepackage{url}
\usepackage{pdfpages}
\usepackage{textcomp}
\usepackage{paralist}

\usepackage[mathlines]{lineno}
\usepackage{lineno}
\newcommand{\bl}{\begin{linenomath}} 
\newcommand{\el}{\end{linenomath}} 

\usepackage{booktabs}
\usepackage{dcolumn}
\newcolumntype{d}[1]{D{.}{.}{#1}}
 
\usepackage{caption}
\newcommand{\real}{\mathbb{R}}

\newcommand{\T}{\mathcal{T}}

\newcommand{\crps}{\textup{crps}}

\newcommand{\cc}{_\textup{c}}

\usepackage{natbib}
\title{Spatial postprocessing of ensemble forecasts for temperature using nonhomogeneous Gaussian regression}

\author{Kira Feldmann\footnote{Ruprecht-Karls-Universit{\"a}t Heidelberg, Germany},\, Michael Scheuerer\footnote{National Oceanic and Atmospheric Administration, Boulder, Colorado, U.S.A.} \,  and  Thordis L. Thorarinsdottir\footnote{Norwegian Computing Center, Oslo, Norway}}

\date{}

\begin{document}

\maketitle

\begin{abstract}
\noindent
Statistical postprocessing techniques are commonly used to improve the skill of ensembles of numerical weather forecasts. This paper considers spatial extensions of the well-established nonhomogeneous Gaussian regression (NGR) postprocessing technique for surface temperature and a recent modification thereof in which the local climatology is included in the regression model for a locally adaptive postprocessing.  In a comparative study employing 21 h forecasts from the COSMO-DE ensemble predictive system over Germany, two approaches for modeling spatial forecast error correlations are considered: A parametric Gaussian random field model and the ensemble copula coupling approach which utilizes the spatial rank correlation structure of the raw ensemble.  Additionally, the NGR methods are compared to both univariate and spatial versions of the ensemble Bayesian model averaging (BMA) postprocessing technique. 
\end{abstract}

\section{Introduction}\label{sec:1}

The first ensemble prediction systems were developed in the early 1990s to account for the various sources of uncertainty in the numerical weather prediction (NWP) model outputs \citep{Lewis2005}. Such systems have now become the state-of-the-art in meteorological forecasting \citep{LeutbecherPalmer2008}.  Additionally, the ensemble forecasts are commonly postprocessed using statistical techniques to improve the calibration and correct for potential biases, and a diverse range of postprocessing techniques has been proposed \citep[e.g.][]{Gneiting2005, Raftery2005, WilksHamill2007, BroeckerSmith2008}.  While these methods have been shown to greatly improve the predictive performance, many are only applicable to univariate weather quantities and neglect forecast error dependencies over time or between different observational sites.  However, correct multivariate dependence structure is often important in applications, especially when considering composite quantities such as minima, maxima or an aggregated 
total. These quantities are crucial e.g. for highway maintenance operations or flood management, where subsequent risk calculations based on the forecast require a calibrated probabilistic forecast for both the original weather variable and the composite quantity. 

In this paper, we focus on spatial extensions of the nonhomogeneous Gaussian regression (NGR) or ensemble model output statistics method for surface temperature, originally proposed by \cite{Gneiting2005}.  NGR is a parsimonious postprocessing technique which, for temperature, returns a Gaussian predictive distribution where the mean value is an affine function of the ensemble member forecasts while the variance is an affine function of the ensemble variance.  The parameters of the model are estimated based on recent forecast errors jointly over a region or separately at each observation location \citep{Gneiting2005, Thorarinsdottir2010}.  Other applications of this approach include \cite{Hagedorn&2008} and \cite{Kann&2009}.  Recently, \cite{ScheuererKoenig2013} proposed a modification of the NGR methods of \cite{Gneiting2005} in which the postprocessing at individual locations varies in space by parameterizing the predictive mean and variance in terms of the local forecast anomalies rather than the forecasts 
themselves. 

To obtain a multivariate predictive distribution based on a deterministic temperature forecast, \cite{Gel&2004} propose the geostatistical output perturbation (GOP) method to generate spatially consistent forecasts fields in which the forecast error field is described through a Gaussian random field model.  \cite{Berrocal2007} combine GOP with the univariate postprocessing method ensemble Bayesian model averaging (BMA) of \cite{Raftery2005}.  Ensemble BMA for temperature dresses each bias corrected ensemble member with a Gaussian kernel and returns a predictive distribution given by a weighted average of the individual kernels. By merging ensemble BMA and GOP, calibrated probabilistic forecasts of entire weather fields are produced.  We propose a similar conceptualization, combining the NGR methods of \cite{Gneiting2005} and \cite{ScheuererKoenig2013} with a Gaussian random field error model in an approach we refer to as spatial NGR.

\begin{figure}
\begin{centering}
\includegraphics[scale=0.33]{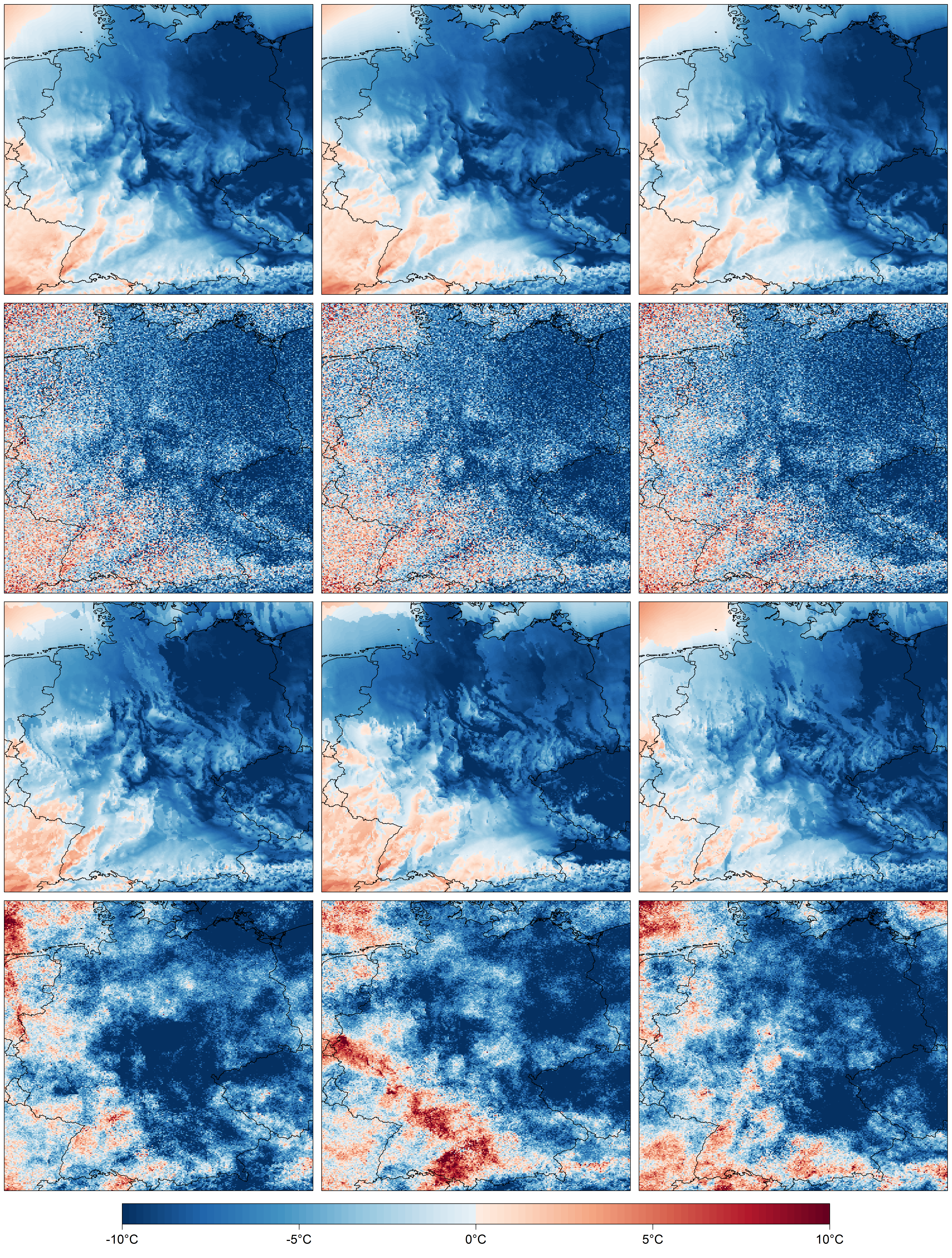}
\par\end{centering}
\caption{21 h temperature forecasts over Germany for January 5, 2011 at 21:00UTC. Three members from the COSMO-DE ensemble prediction system are shown in the top row, the second row demonstrates the postprocessed nonhomogeneous Gaussian regression (NGR) forecast, the third row shows NGR combined with ensemble copula coupling while the bottom row displays examples of the multivariate spatial NGR forecast.}
\label{fig:tempfields}
\end{figure}

As an alternative multivariate method, we consider the non-parametric ensemble copula coupling (ECC) approach of \cite{Schefzik&2013}. ECC returns a postprocessed ensemble of the same size as the original raw ensemble.  The prediction values at each location are samples from the univariate postprocessed predictive distribution at that location. Multivariate forecast fields are subsequently generated using the rank correlation structure of the raw ensemble. ECC thus assumes that the ensemble prediction system correctly describes the spatial dependence structure of the weather quantity. The method applies equally to any multivariate setting and comes at a virtually no additional computational cost once the univariate postprocessed predictive distributions are available.  

Figure~\ref{fig:tempfields} illustrates temperature field forecasts obtained from the raw ensemble, the standard univariate NGR method, NGR combined with ECC, and spatial NGR.  The raw ensemble is depicted in the first row. The NWP model output has a physically consistent spatial structure, but as we shall see later, it is strongly underdispersive and does not adequately represent the true forecast uncertainty. The samples in rows 2-4 all share the same NGR marginal predictive distributions which have larger uncertainty bounds than the raw ensemble. In the second row, the realizations have been sampled independently for each gridpoint i.e.\ no spatial dependence structure is present.  This results in unrealistic temperature fields and, when considering compound quantities, forecasts that are statistically inappropriate. The combination of NGR and ECC in the third row gives forecast fields with similar spatial structures as the raw ensemble even though there is larger spread both within each field and 
between the realized fields.  As a consequence of spatial correlations being modeled through a discrete copula, the resulting temperature fields feature some sharp transitions at locations where the ranks of the raw ensemble change. The bottom row depicts temperature field simulations obtained with spatial NGR.  Here, the spatial dependence between forecast errors at different locations is modeled by a statistical correlation model and the physical consistency is implicitly learned from the data.

In a comparative study, we apply the various extensions of NGR as well as ensemble BMA to 21 h forecasts of surface temperature over Germany issued by the German Weather Service through their COSME-DE ensemble prediction system. The remainder of the paper is organized as follows. The forecast and observation data are described in the next Section 2. The univariate NGR postprocessing methods are introduced in Section 3 while the multivariate methods are described in Section 4. In the following Section 5 we report the results of the case study and a discussion is provided in Section 6.  Finally, an overview over the forecast evaluation methods applied in Section 5 is given in the Appendix. 

\section{Forecast and observation data}\label{sec:data}

\begin{figure}
\begin{centering}
\includegraphics[scale=0.4]{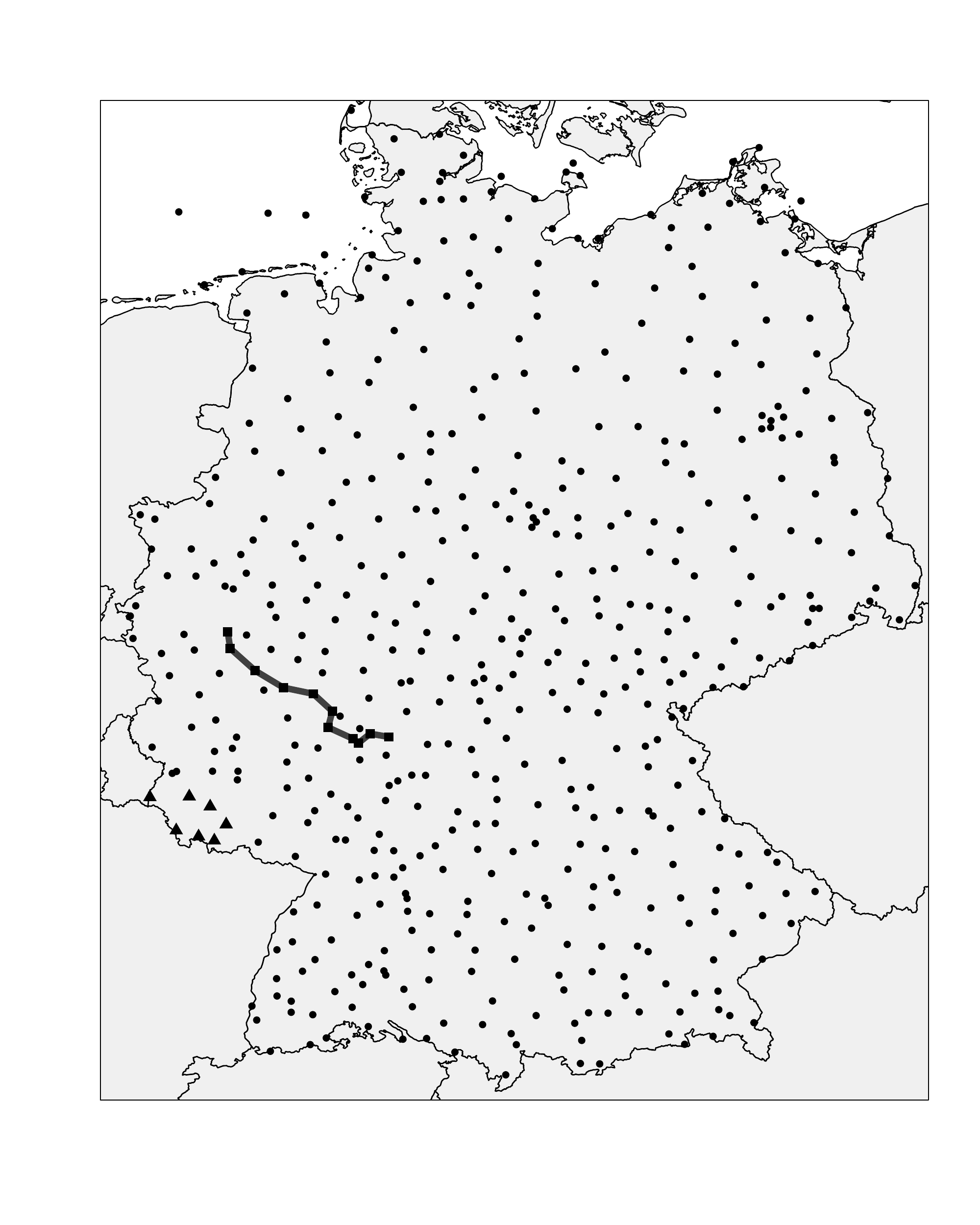}
\par\end{centering}
\caption{Map of Germany showing the location of a total of 514 SYNOP observation stations. The gray line illustrates a section of the highway A3 with the 
surrounding observation stations indicated by solid squares. Seven observation stations in the state of Saarland are represented by solid triangles. Other stations are indicated by solid circles.}
\label{fig:stations}
\end{figure}

The COSMO-DE forecast dataset consists of a 20-member ensemble.  Forecasts are made for lead times from 0 to 21 h on a 2.8 km grid covering Germany with a new model run being started every 3 h.  The ensemble is based on a convection-permitting configuration of the NWP model COSMO \citep{Steppeler2003, Baldauf2011}.  It has a $5 \times 4$ factorial design with 5 different perturbations in the model physics and 4 different initial and boundary conditions provided by global forecasting models \citep{Gebhardt2011, Peralta2011}.  The pre-operational phase of the COSMO-DE ensemble prediction system started on 9 December 2010 and the operational phase was launched on 22 May 2012.

We employ 21 h forecasts from the pre-operational phase initialized at 00:00 UTC; our entire dataset consists of forecasts from 10 December 2010 until 30 November 2011.  As we use a rolling training period of 25 days to fit the parameters of the statistical postprocessing methods, the evaluation period runs from 5 January 2011 to 30 November 2011.  If at least one ensemble member forecast is missing at all observation locations on a specific day, we omit this day from the dataset. This way, 10 days are eliminated with 346 days remaining.  The temperature observations we employ stem from 514 SYNOP stations over Germany. Their locations are shown in Figure~\ref{fig:stations}.  The forecasts are interpolated from the forecast grid to the station locations using bilinear interpolation.  In total, we evaluate forecasts for 117,879 verifying observations over 346 days. The COSMO model uses a rotated spherical coordinate system in order to project the geographical coordinates to the plane with distortions as small 
as possible \citep[Sec.~3.3]{DomsSchaettler2002}, with $421\times461$ equidistant gridpoints in longitudinal and latitudinal direction. We adopt this coordinate system to calculate horizontal distances within the framework of our spatial correlation models.

\section{Univariate postprocessing}\label{sec:3}

\subsection{NGR for temperature}\label{subsec:EMOS}

The NGR method of \cite{Gneiting2005} generalizes the common model output statistics (MOS) postprocessing technique, see e.g.~\cite{Wilks2011}. The distribution of the future state $y_{s}$ of the temperature at location $s$ is modeled as a Gaussian distribution with parameters depending on the $M$ ensemble forecasts $f_{1s},...,f_{Ms}$,
\bl
\begin{equation}\label{eq:NGR}
y_s|f_{1s},...,f_{Ms}\sim\mathcal{N}(a+b_1f_{1s}+\ldots+b_Mf_{Ms},c+dS_s^2),
\end{equation}
\el
where $S_s^2$ is the ensemble variance, $a,b_1,\ldots,b_M \in \real$ are regression coefficients and $c,d \in \real_+$ are non-negative coefficients.   NGR is thus a linear model with the ensemble forecasts as predictors and a nonhomogeneous error term which is modeled as an affine function of the ensemble variance $S_{s}^{2}$. This modeling set-up counteracts a possible over- or underdispersion of the ensemble while exploiting a positive spread-error correlation. The normal predictive distribution presents a reasonable model for variables like temperature or surface pressure.  While the Gaussian assumption is not appropriate for all weather variables, the basic idea of NGR can also be used with other types of predictive distributions \citep{Thorarinsdottir2010, Thorarinsdottir2012, LerchThorarinsdottir2013, Scheuerer2013}. 

In the formulation in \eqref{eq:NGR}, the regression coefficients $b_{1},...,b_{M}$ can take any value in $\real$. However, as negative values are difficult to interpret, \citet{Gneiting2005} suggest an alternative formulation restricting the coefficients $b_1, \ldots, b_M$ to be non-negative by iteratively removing those ensemble members $f_m$ from the linear model for which the coefficients $b_m$ are negative.  We follow \cite{Thorarinsdottir2010} and obtain non-negative coefficients by setting $b_1 = \beta_1^2, \ldots, b_M = \beta_M^2$ with $\beta_1, \ldots, \beta_M \in \real$.  The (normalized) coefficients can then be interpreted as weights and reflect the relative performance of the ensemble members during the training period. In the following, we refer to this approach as NGR$_+$. 

\subsection{Locally adaptive NGR}\label{subsec:AdaptiveEMOS}

The NGR postprocessing in \eqref{eq:NGR} makes the same adjustments of ensemble mean and variance at all locations.  However, it has been argued that systematic model biases may vary in space due to e.g. incomplete resolution of the orography or different land use characteristics.  Similarly, the prediction uncertainty may differ between locations in a way that is not represented by the ensemble spread \citep{Kleiber2011, ScheuererBuermann2013,ScheuererKoenig2013}.  

As an alternative to the NGR model in \eqref{eq:NGR}, we consider the locally adaptive NGR method of \cite{ScheuererKoenig2013}.  Here, the local adaptation is obtained by incorporating information about the short-term local climatology in the covariates of the predictive distribution, thereby using forecast anomalies rather then the original forecasts as covariates. Specifically, let $\bar{y}_s^{\T}$ denote the average observed local temperature at location $s$ over all days $t$ in the training period $\T$ and, correspondingly, denote by $\bar{f}_{ms}^{\T}$ the average temperature forecast by the $m$-th ensemble member. The predictive distribution then equals 
\bl
\begin{equation}\label{eq:NGRc}
y_s|f_{1s},...,f_{Ms},\mathcal{D}_s^{\T} \sim \mathcal{N}\big(\bar{y}_s^{\T} + b_1(f_{1st}-\bar{f}_{1s}^{\T}) + \ldots + b_M(f_{Ms}-\bar{f}_{Ms}^{\T}), c\xi_s^2 + dS_{s}^2\big),
\end{equation}
\el
where $\mathcal{D}_s^{\T}$ denotes the forecast and observation data at location $s$ during the training period and $\xi_s^2$ is a predictor variable for the location specific uncertainty.  The location specific uncertainty is defined in a second step, when the regression estimates $\hat{b}_1,\ldots,\hat{b}_M$ are available. It is given by the mean squared residuals of the regression fit,
\[
\xi_s^2 = \frac{1}{|\T|} \sum_{t \in \T} \bigg( y_{st} - \bar{y}_s^{\T} - \sum_{m=1}^M \hat{b}_m \big(f_{mst} - \bar{f}_{ms}^{\T}\big)\bigg)^2, 
\]
where $| \T |$ denotes the number of days in the training period.  At locations without an observation station, predictive means and variances are obtained by spatial interpolation as described in \cite{ScheuererKoenig2013}. We refer to this inclusion of the local climatology as NGR$_\textup{c}$. 

\subsection{Parameter estimation in the univariate setting}

We focus on the NGR$_+$ formulation of the NGR method in \eqref{eq:NGR} and the NGR$_\textup{c}$ extension in \eqref{eq:NGRc}. The parameter estimation for both methods proceeds in a similar manner.  It is assumed that the forecast error statistics change only slowly over time and a rolling training window $\T$ of the $| \T |$ most recent dates with forecasts and observations available is used to estimate the model parameter.  The model fitting is carried out with all available data from the set of observation sites $\mathcal{S}$ within the training window $\T$. We denote this set by $\mathcal{S}$ in the equations below even though observations might not always be available at all observation sites.  Note that with NGR$_+$ we can easily generate postprocessed forecasts outside the set $\mathcal{S}$ since the parameters of the postprocessing techniques are not site-specific. When NGR$_\textup{c}$ is used, this can be achieved through an additional spatial interpolation step.

We follow \citet{Gneiting2005} and estimate the NGR$_+$ parameters by minimizing the continuous ranked probability score (CRPS) \citep[e.g.][]{Gneiting2007} over the training set. That is, we chose them as a solution to
\bl
\begin{equation}\label{Eq:MeanCRPS}
\min_{a,b_1,\ldots,b_M,c,d}
 \frac{1}{|\T| |\mathcal{S}|}\sum_{t\in\mathcal{T}}\sum_{s\in\mathcal{S}} \crps(\Phi_{st},y_{st}), 
\end{equation}
\el
where $\Phi_{st}$ is the Gaussian distribution function in \eqref{eq:NGR} on training day $t$ at site $s$, $y_{st}$ is the corresponding verifying observation and 
\bl
\begin{equation}\label{eq:crps}
 \crps(\Phi_{st},y_{st}) = \int_{-\infty}^\infty \big(\Phi_{st}(x)-{\bf 1}_{[y_{st},\infty)}(x)\big)^2 dx,
\end{equation}
\el
with ${\bf 1}_{[y_{st},\infty)}(x)$ equal to 1 if $x \in [y_{st},\infty)$ and 0 otherwise. For Gaussian distributions, the integral in \eqref{eq:crps} can be expressed in a closed form which minimizes the computational costs \citep{Gneiting2005}. Software for the estimation and prediction is available through the {\tt ensembleMOS} package in {\tt R} \citep{R2013} which is available at \url{www.r-project.org}.

The parameters of the NGR$\cc$ method, on the other hand, are estimated in two steps. In a first step, the regression parameters $b_1,\ldots,b_M$ in \eqref{eq:NGRc} are estimated by weighted least squares using a penalized version of the loss function to prevent overfitting, see \cite{ScheuererKoenig2013} for details.  The parameters of the variance function, $c$ and $d$, are subsequently estimated via CRPS minimization as in \eqref{Eq:MeanCRPS} above. 

\section{Multivariate methods}\label{sec:4}

\subsection{Ensemble copula coupling} \label{subsec:ECC}

The ensemble copula coupling (ECC) method of \cite{Schefzik&2013} employs the rank order structure of the raw ensemble to obtain a postprocessed ensemble of forecasts fields with the same multivariate correlation structure as the raw ensemble while retaining the univariate NGR marginals. It is thus a semi-parametric copula approach with continuous marginals and the non-parametric empirical copula.  For each $s \in \mathcal{S}$, we draw a sample of size $M$ from the predictive distribution $\hat{\Phi}_s$ given by \eqref{eq:NGR} or \eqref{eq:NGRc} of the form
\bl
\begin{equation}\label{eq:ECCsample}
\hat{f}_{1s} = \hat{\Phi}_s^{-1}\Big( \frac{1}{M+1}\Big), \ldots, \hat{f}_{Ms} = \hat{\Phi}_s^{-1}\Big( \frac{M}{M+1}\Big).
\end{equation}
\el
That is, it holds that $\hat{f}_{1s} \leq \cdots \leq \hat{f}_{Ms}$.  Let $\rho_s$ denote a permutation of the integers $\{1, \ldots, M\}$ defined by $\rho_s(m) = \textup{rank}(f_{ms})$ for $m = 1, \ldots, M$ with ties resolved at random.  Then it follows that the sample $\{ \hat{f}_{\rho_s(1)s}, \ldots, \hat{f}_{\rho_s(M)s}\}$ has the same rank order structure as the raw ensemble $\{f_{1s}, \ldots, f_{Ms}\}$.  The ECC ensemble of postprocessed forecast fields is thus given by  
\bl
\begin{equation}\label{eq:ECC}
\{ \hat{f}_{\rho_s(m)s}\}_{s \in \mathcal{S}}, \quad m = 1, \ldots, M.
\end{equation}
\el
\cite{Schefzik&2013} discuss and compare several alternative methods to draw the sample from each univariate predictive distribution. We use the quantiles in \eqref{eq:ECCsample} as it follows from results in \cite{Broecker2012} that this sample maintains the calibration of the univariate forecasts.

\subsection{Spatial NGR}\label{subsec:SpatialEMOS}

The geostatistical output perturbation \citetext{GOP, \citealp{Gel&2004}} approach was originally introduced as an inexpensive substitute of a dynamical ensemble based on a single numerical weather prediction. It dresses the deterministic forecast with a simulated forecast error field according to a spatial random process, thus perturbing the outputs of the NWP models rather than their inputs.  We propose a spatial NGR method that adopts the ideas from GOP and combines them with the univariate NGR methods described in Sections \ref{subsec:EMOS} and \ref{subsec:AdaptiveEMOS}.  The result is a multivariate predictive distribution which generates spatially coherent forecast fields while retaining the univariate NGR marginals.  The spatial NGR method can thus also be seen as a fully parametric Gaussian copula approach \citep{Moeller&2013, Schefzik&2013}.

Denote by $\mathbf{Y}=\left\{ y_{s}:\: s\in\mathcal{S}\right\}$ the vector whose components represent the temperature at each location in $\mathcal{S}$, and by $\mathbf{F}_m=\left\{ f_{ms}:\: s\in\mathcal{S}\right\} $ the corresponding weather field forecast by the $m$-th ensemble member.  The vector $\boldsymbol{\mu}$ of predictive means obtained by marginal NGR postprocessing is given by 
\begin{equation}\label{eq:spatialNGRmean}
 \boldsymbol{\mu} = \left\{
 \begin{array}{ll}
  a\boldsymbol{1} + b_1\mathbf{F}_1 + \ldots + b_M\mathbf{F}_M & (\mbox{NGR}_+), \\
  \overline{\mathbf{Y}}^{\mathcal{T}} + b_1\big(\mathbf{F}_1-\overline{\mathbf{F}}_1^{\mathcal{T}}\big) + \ldots + b_M\big(\mathbf{F}_M-\overline{\mathbf{F}}_M^{\mathcal{T}}\big) & (\mbox{NGR}_c),
 \end{array}
 \right.
\end{equation}
where $\boldsymbol{1}$ is a vector of length $|\mathcal{S}|$ with all entries equal to $1$, $\overline{\mathbf{Y}}^{\mathcal{T}}$ is the average observed temperature vector over the training period $\mathcal{T}$ and  $\overline{\mathbf{F}}_m^{\mathcal{T}}$ denotes the vector consisting of the average temperature forecast by the $m$-th ensemble member over $\mathcal{T}$, see also \eqref{eq:NGRc}.  Similarly, denote by $\mathbf{D}$ the diagonal matrix of the univariate predictive standard deviations $\sigma_s$ with $\sigma_s^2=c+dS_s^2\:$ for NGR$_+$ and $\sigma_s^2=c\xi_s^2+dS_s^2\:$ for NGR$_c$.

The spatial NGR multivariate predictive distribution corresponds to the sum of the bias-corrected forecast mean vector given by \eqref{eq:spatialNGRmean}, a zero-mean random vector with correlated components, and a further zero-mean random vector with uncorrelated components representing small-scale variations that cannot be resolved with the available data. That is,
\bl
\[\mathbf{Y}|\mathbf{F}_1,\ldots,\mathbf{F}_M,\mathcal{D}_{\mathcal{S}}^{\mathcal{T}} = \boldsymbol{\mu} + \mathbf{D} \tilde{\mathbf{E}}
\]
\el
where 
\bl
\[
\tilde{\mathbf{E}} = \sqrt{1-\theta}\cdot\mathbf{E}_1 + \sqrt{\theta}\cdot\mathbf{E}_2, \qquad \theta \in [0,1]. 
\]
\el
If all components of $\mathbf{E}_1$ and $\mathbf{E}_2$ have unit variance, the multiplication with $\mathbf{D}$ scales the components of $\tilde{\mathbf{E}}$ such that their variances match those predicted by the univariate NGR postprocessing methods.  In particular, the resulting multivariate model features spatially varying predictive variances.  

For the spatial correlations we follow \citet{Gel&2004} and assume a stationary and isotropic correlation function $C_{\theta,r}$ of the exponential type.  That is, we assume that the correlation between two components of $\tilde{\mathbf{E}}$ corresponding to locations $s_i$ and $s_j$ depends only on their Euclidean distance $\|s_i-s_j\|$ and is given by
\bl
\begin{equation}\label{eq:spatialNGRcov}
C_{\theta,r}\left(s_i,s_j\right) = (1-\theta)\cdot e^{-\frac{\|s_i-s_j\|}{r}} + \theta\cdot \delta_{ij},
\end{equation}
\el
where $\delta_{ij}$ denotes to the Kronecker delta function which is equal to $1$ if $i = j$ and $0$ otherwise.  The parameter $\theta\in[0,1]$ has already been introduced above and controls the relative contribution of the spatially correlated random vector $\mathbf{E}_1$ and the spatially uncorrelated random vector $\mathbf{E}_2$ to the overall variance. The range parameter $r>0$ determines the rate at which the spatial correlations of $\mathbf{E}_1$ decay with distance.  Once those parameters have been estimated, samples of $\tilde{\mathbf{E}}$ can be simulated, scaled by the site specific standard deviations, and added to the forecast mean vector $\boldsymbol{\mu}$.  

The resulting spatial NGR multivariate predictive distribution at locations within $\mathcal{S}$ is given by 
\bl
\begin{equation}\label{eq:spatialNGR}
\mathbf{Y}|\mathbf{F}_1,\ldots,\mathbf{F}_M,\mathcal{D}_{\mathcal{S}} \sim \mathcal{N}_{| \mathcal{S} |}\left( \boldsymbol{\mu}, \boldsymbol{\Sigma}\right), \quad \mbox{ with } \; \boldsymbol{\Sigma} = \mathbf{D}\mathbf{P}\mathbf{D},
\end{equation}
\el
where $\mathcal{N}_{| \mathcal{S} |}$ denotes a multivariate normal distribution of dimension $|\mathcal{S}|$, and $\mathbf{P}$ is the correlation matrix of $\tilde{\mathbf{E}}$.  Note that this definition can be extended to locations outside the set $\mathcal{S}$; as pointed out above, both $\boldsymbol{\mu}$ and $\mathbf{D}$ can be defined for any set of locations where ensemble forecasts are available, and the same is also true for $\mathbf{P}$ since (\ref{eq:spatialNGRcov}) presents a well-defined correlation function over the entire Euclidean plane.

\subsection{Estimating the spatial NGR correlation parameters}\label{subsec:ParamEst}

In order to estimate the correlation parameters $\theta$ and $r$ in \eqref{eq:spatialNGRcov}, we consider the standardized forecast errors $\tilde{e}_{st} := (y_{st}-\mu_{st})/\sigma_{st}$ at all locations $s\in\mathcal{S}$ and on all training days $t\in\mathcal{T}$, and study their average half squared differences over $\mathcal{T}$,
\bl
\begin{equation}\label{Eq:AvgHalfSqDiff}
 \frac{1}{2|\mathcal{T}|}\sum_{t\in\mathcal{T}}\big(\tilde{e}_{s_it}-\tilde{e}_{s_jt}\big)^2, \qquad s_i,s_j\in\mathcal{S}.
\end{equation}
\el
This way we obtain an empirical version of the variogram
\bl
\begin{equation}\label{eq:spatialNGRvg}
\gamma_{\theta,r}(s_i,s_j) := \frac{1}{2} \mathrm{Var}\big(\tilde{\mathbf{E}}_{s_i}-\tilde{\mathbf{E}}_{s_j}\big) = (1-\theta)\left(1-e^{-\frac{\|s_i-s_j\|}{r}}\right) + \theta(1-\delta_{ij}),
\end{equation}
\el
that corresponds to our correlation model $C_{\theta,r}$. 

The assumption of a stationary and isotropic correlation function implies that $\gamma_{\theta,r}(s_i,s_j)$ is a function of the distance $d(s_i,s_j)=\|s_i-s_j\|$ only, and so the average half squared differences in \eqref{Eq:AvgHalfSqDiff} can be further aggregated which reduces their variability. Specifically, we sort all pairs $s_i,s_j$ by their distance into bins $B_1,\ldots,B_L$ with the bin sizes chosen such that the number of pairs in each bin is approximately equal. The values in \eqref{Eq:AvgHalfSqDiff} are then averaged over each bin, resulting in pairs $(d_l,\hat{\gamma}_l)_{l=1,\ldots,L}$ of distances associated with each bin and empirical variogram values that may be compared to the theoretical variogram (\ref{eq:spatialNGRvg}) for given values of $\theta$ and $r$. For the calculation of these values, we employ the $\mathtt{R}$ package $\mathtt{RandomFields}$ by \citet{Schlather2011}. When fitting a curve to the empirical variogram, we 
follow \citet{Berrocal2007} and use weighted 
least squares fitting as proposed by \citet{Cressie1985}, minimizing the function 
\bl
\[ S(\theta,r) = \sum_{l=1}^L n_l \left(\frac{\hat{\gamma}_l-\gamma_{\theta,r}(d_l)}{\gamma_{\theta,r}(d_l)}\right)^2, \]
\el
where $n_{l}$ is the number of location pairs associated with $B_{l}$. The minimization problem has to be solved numerically for which we use the optimization algorithm by \citet{Bryd1995} as implemented in the {\tt R}-function $\mathtt{optim}$ \citep{R2013}.  The range parameter $r$ is constrained to be positive and not larger than the maximum distance over the entire domain, which equals 890 km. The starting values are kept fixed at the averaged values over the entire forecasting period obtained in earlier experiments.

Alternatively, the random field parameters could be estimated via maximum likelihood. Under ideal conditions, this is statistically more efficient. However, as maximum likelihood estimation is potentially more sensitive to outliers and computationally more expensive we employ the variogram based approach in line with \cite{Berrocal2007}. Indeed, results obtained with maximum likelihood estimation (not shown here) slightly reduced the predictive performance of the spatial NGR forecasting methods.

\section{Results}\label{sec:5}

In this section we present the results of applying the univariate NGR$_+$ and NGR$_{\textup c}$ postprocessing methods as well as their spatial extensions to forecasts from the COSME-DE ensemble prediction system, described in Section~\ref{sec:data}.  Additionally, we provide a comparison to the univariate ensemble BMA method of \cite{Raftery2005} and the multivariate spatial BMA approach, proposed by \cite{Berrocal2007}. The forecast evaluation methods are discussed in the Appendix. 

\subsection{Ensemble BMA reference methods}

The univariate ensemble BMA method of \cite{Raftery2005} is a kernel dressing approach where, for temperature, each bias corrected ensemble member is dressed with a Gaussian kernel, where the variance is kept fixed. That is, 
\bl
\[
y_s | f_{ms} \sim \mathcal{N} ( a_m + b_m f_{ms} , \sigma^2),
 \]
\el
for $m = 1, \ldots, M$ and $s \in \mathcal{S}$.  The predictive density is then given by a weighted average
\bl
\begin{equation}\label{eq:BMA}
\sum_{m=1}^M \omega_m \varphi(a_m + b_m f_{ms}, \sigma^2), 
\end{equation}
\el
where $\varphi$ denotes the Gaussian density and the weights $\omega_1, \ldots, \omega_M$ are nonnegative with $\sum_{m=1}^M \omega_m = 1$. The weights reflect the skill of each ensemble member in the training period $\mathcal{T}$.  We estimate the ensemble BMA parameters using the {\tt R} package {\tt ensembleBMA} employing the same training period $\mathcal{T}$ as for the univariate NGR methods. 

The spatial BMA approach of \cite{Berrocal2007} combines ensemble BMA with the GOP method of \cite{Gel&2004} in a similar way as the spatial NGR methods, described in Section~\ref{subsec:SpatialEMOS} above.  However, it differs from spatial NGR in the manner in which realizations of the multivariate predictive distribution are simulated.  For a temperature forecast field under spatial BMA, we first randomly choose a member of the dynamical ensemble according to the ensemble BMA weights in \eqref{eq:BMA}, and then dress the corresponding bias-corrected forecast field with an error field that has a stationary covariance structure specific to this member.  As the forecast field is chosen at random and the covariance function is member-specific, the final covariance structure becomes non-stationary. This comes at the expense of having to estimate $M$ different covariance functions, while spatial NGR requires a single correlation function only, and achieves spatially varying variances through scaling. 

\subsection{Univariate predictive performance}

Measures of univariate predictive performance of the raw COSMO-DE ensemble and the postprocessed forecasts under NGR$_+$, NGR$_{\textup c}$, and BMA are given in Table~\ref{tab:univ results}.  A simple approach to assess calibration and sharpness of univariate probabilistic forecasts is to calculate the nominal coverage and width of prediction intervals.  The 90.5\% prediction interval considered here corresponds to the probability that the observation is within the ensemble range, assuming that the ensemble members and the observation are exchangeable. While the raw ensemble returns very sharp forecasts, it is severely underdispersive as can be seen by the insufficient coverage.  This is also reflected in the numerical scores which are significantly better for all three postprocessing methods. NGR$_+$ and ensemble BMA return essentially identical scores, improving upon the ensemble by $34$\% in terms of the CRPS and by approximately $18$\% in terms of both mean absolute error (MAE) and root mean squared 
error (RMSE).  Ensemble BMA returns minimally wider prediction intervals than NGR$_+$ but yields an empirical coverage that is closest to the nominal 90.5\%.  The locally adaptive postprocessing of NGR$_{\textup c}$ yields the best overall scores and approximately $10$\% shorter prediction intervals than NGR$_+$ on average despite being slightly underdispersive.  The station-specific reliability indices indicate that the postprocessing improves the calibration consistently across the country with the postprocessing methods always yielding lower indicies than the average of the ensemble. 

\begin{table}[!bthp]
  \centering
  \caption{Mean continuous ranked probability score (CRPS), mean absolute error (MAE) and root mean squared error (RMSE) for 21 h temperature forecasts aggregated over all 514 stations and 346 days in the test set. Also reported here are the average width (PI-W) and coverage (PI-C) of 90.5\% prediction intervals aggregated over the entire test set and the mean (RI-Mean), minimum (RI-Min) and maximum (RI-Max) station-specific reliability indicies. }\label{tab:univ results}
  \smallskip
  \begin{tabular}{lcccccccc}
    \toprule 
    & CRPS & MAE & RMSE & PI-W & PI-C & RI-Mean & RI-Min & RI-Max \\
    & (\textcelsius)   & (\textcelsius) & (\textcelsius) & (\textcelsius) & (\%) & \\
    \midrule 
    Raw ensemble       & 1.56  & 1.77  & 2.27  & 1.50  & 26.0  & 1.29 & 0.72 & 1.63\\
    BMA                & 1.04  & 1.46  & 1.86  & 5.91  & 88.9  & 0.35 & 0.14 & 0.97\\
    NGR$_+$            & 1.04  & 1.46  & 1.87  & 5.76  & 88.0  & 0.35 & 0.12 & 0.93\\
    NGR$_{\textup c}$  & 0.96  & 1.35  & 1.73  & 5.17  & 86.6  & 0.23 & 0.13 & 0.84\\
    \bottomrule  
  \end{tabular}
\end{table}

\subsection{Spatial calibration}

\begin{figure}
\begin{centering}
\includegraphics[scale=0.9]{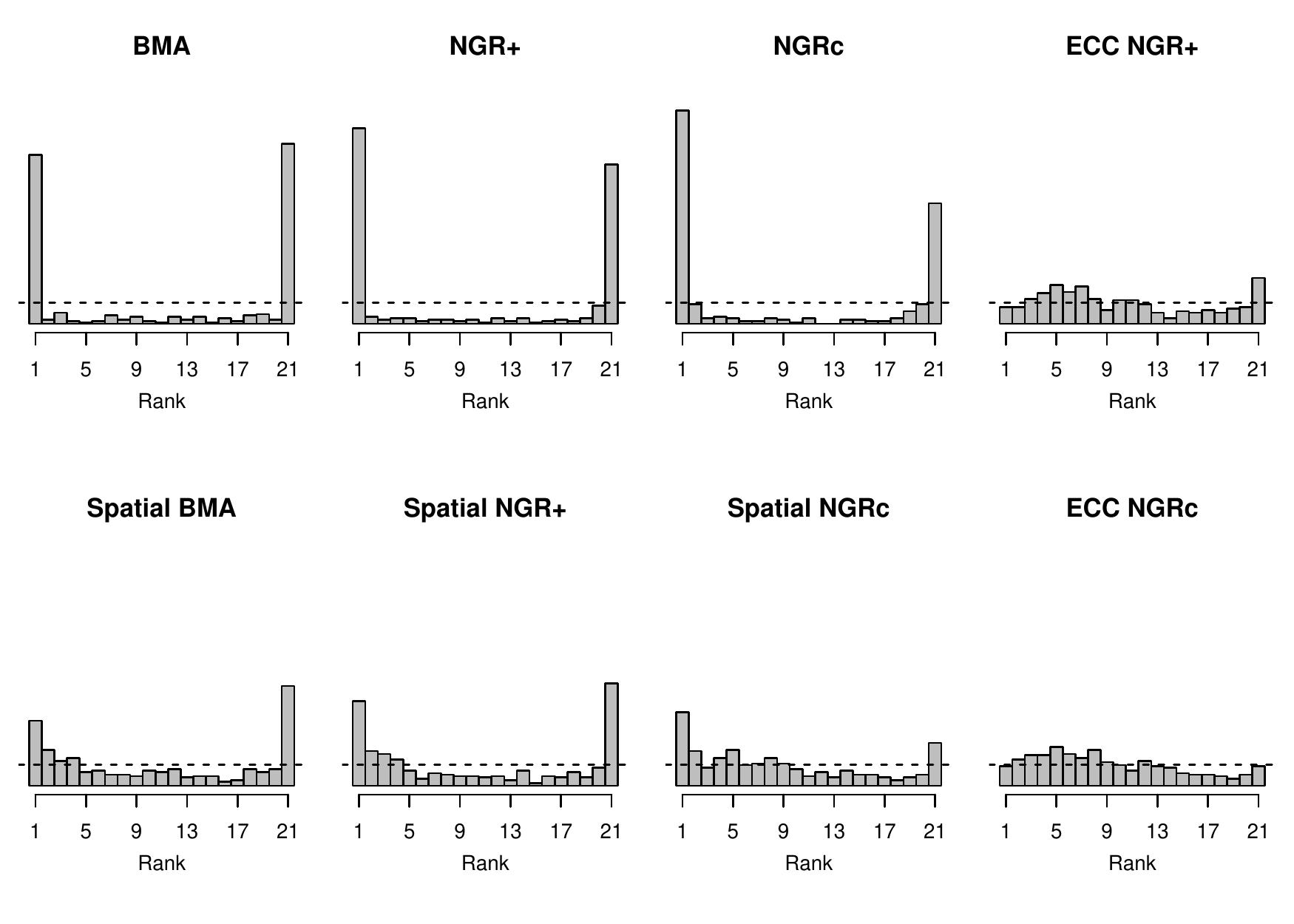}
\par\end{centering}
\caption{Multivariate band depth rank histograms to assess the calibration of joint forecast fields at 514 stations in Germany aggregated over the 346 days in the test set.}
\label{fig:arGer}
\end{figure}

In Figure~\ref{fig:arGer} we assess the calibration of the joint forecast fields at all 514 observation stations in Germany using multivariate band depth rank histograms (see the Appendix for details).  Without additional spatial modeling (i.e.\ assuming independent forecast errors at the different stations) the multivariate calibration of BMA, NGR$_+$, and NGR$_{\textup c}$ is rather poor, despite their good marginal calibration.  The three spatial forecasts that are based on parametric modeling of the error field -- spatial BMA, spatial NGR$_+$ and spatial NGR$_{\textup c}$ -- significantly improve upon the calibration of the univariate methods, in particular spatial NGR$_\textup{c}$.  However, the strength of the correlations seems somewhat too low as the observed field is too often either the most central or the most outlying field resulting in a $\cup$-shaped histogram \citetext{see also Section 4 of \citealp{Thorarinsdottir&2013b}}. In contrast, the combination of ECC and NGR produces forecasts fields where the strength of the correlations appears slightly too high. This result is supported by the spatial correlation patterns portrayed in Figure~\ref{fig:tempfields} where the raw ensemble -- and thus also the ECC fields -- appears to have significantly longer-range correlations than the estimated Gaussian error fields.     

\begin{figure}
\begin{centering}
\includegraphics[scale=0.55]{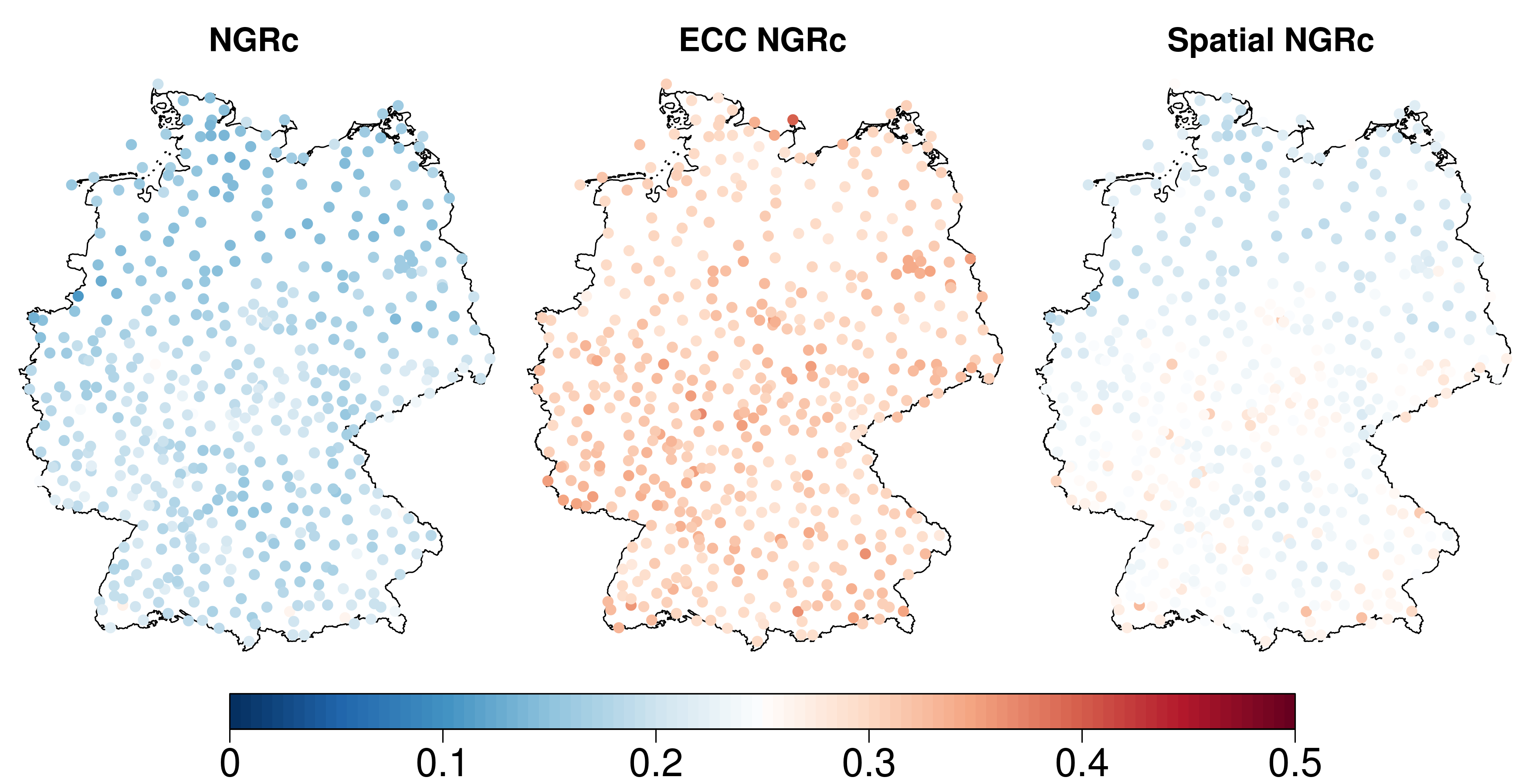}
\par\end{centering}
\caption{Mean absolute deviations of the temperature difference PIT values from $0.5$ where the mean is taken over all verification days and, for each station, over all stations within a radius of $50$ km.}
\label{fig:differencePITs}
\end{figure}

In our spatial NGR$_+$/NGR$_{\textup c}$ model we made the simplifying assumption of a stationary and isotropic correlation function.  In order to check whether this assumption is appropriate or whether correlation strengths vary strongly over the domain considered here, we study the probability integral transforms (PITs) of predicted temperature differences between close-by stations (a motivation and interpretation for that is given in the Appendix).  Specifically, we focus on the NGR$_{\textup c}$ model and its spatial extensions where we can assume that the univariate predictive distributions have no local biases and reflect the local prediction uncertainty reasonably well \citep{ScheuererKoenig2013}.  Figure~\ref{fig:differencePITs} depicts, for each station, the mean absolute deviations of the PIT values from $0.5$ over all verification days and all temperature differences between this station and stations within a $50$ km neighborhood.  As expected, in the absence of a spatial model the magnitude of temperature differences is overestimated.  When ECC is used to restore the rank correlations of the raw ensemble, it is underestimated (i.e.\ spatial correlations are too strong), which is in line with our conclusions from Figure~\ref{fig:arGer}.  On average, the mean absolute deviations from $0.5$ of the PIT values corresponding to spatial NGR$_{\textup c}$ are closest to the value $0.25$ that corresponds to perfect calibration.  However, the adequate correlation strength varies across the domain.  The assumption of stationarity and isotropy of our statistical correlation model entails too weak correlations over the North German Plain and too strong correlations near the Alpine foothills and in the vicinity of the various low mountain ranges.  After all, (\ref{eq:spatialNGRcov}) presents a good first approximation, but a more sophisticated, non-stationary correlation model may yield further improvement.

\subsection{Case study I: Predictive performance in Saarland}

\begin{figure}
\begin{centering}
\includegraphics[scale=0.9]{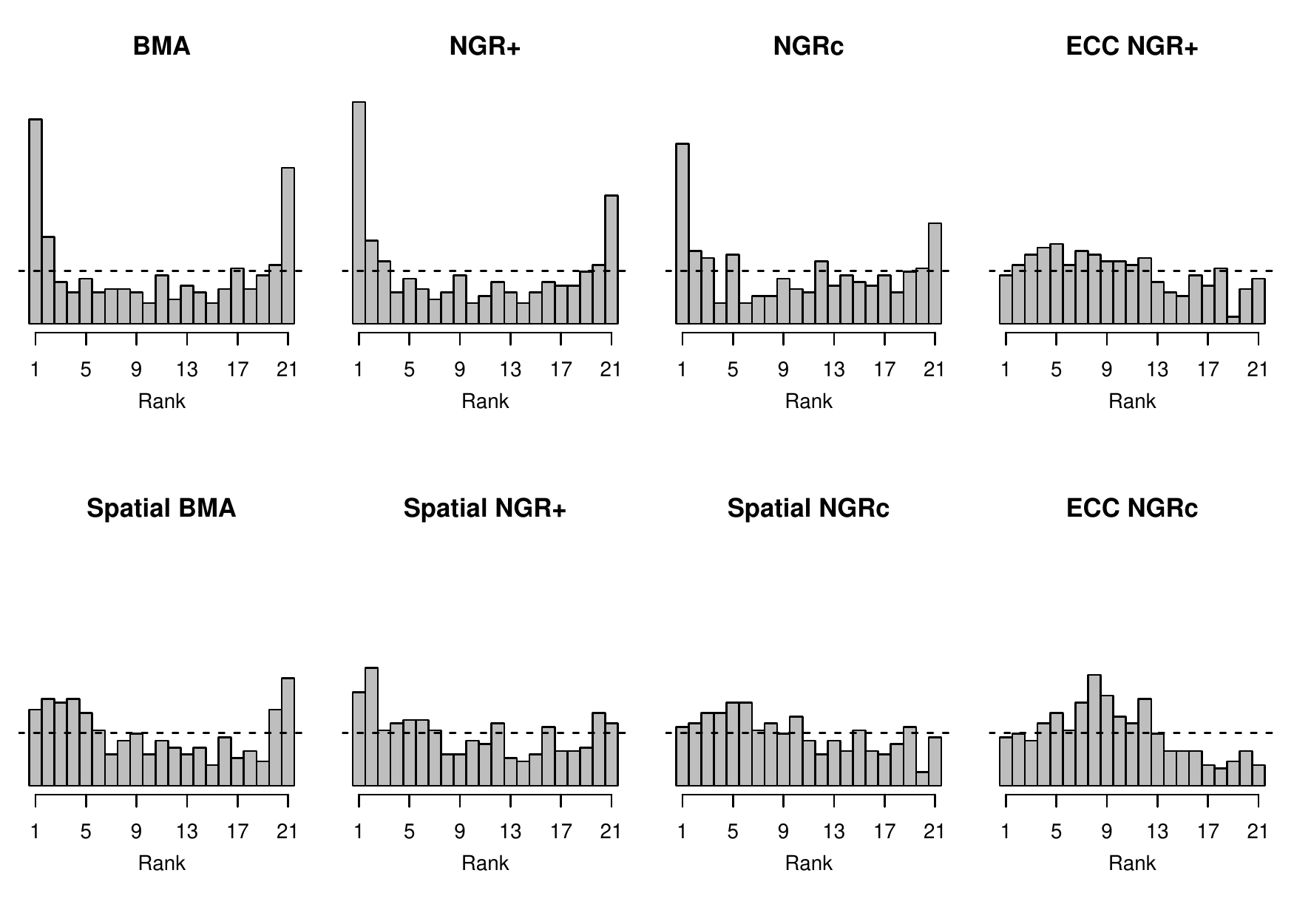}
\par\end{centering}
\caption{Multivariate band depth rank histograms to assess the calibration of joint temperature forecasts at the seven observation stations in the state of Saarland in Germany aggregated over the 346 days in the test set.}
\label{fig:sar}
\end{figure}

For a more quantitative assessment of multivariate predictive performance, we focus on two smaller subsets of the 514 stations.  This is necessary because in our own experience, the lack of sensitivity of the energy score to misspecifications of the spatial correlation structure \citep{PinsonTastu2013} becomes even worse as the dimension of locations considered simultaneously increases.

First, we  consider the joint predictive distribution at the seven stations in the state of Saarland (see Figure~\ref{fig:stations}).  The corresponding multivariate band depth rank histograms in Figure~\ref{fig:sar} confirm the conclusions from the preceding subsection in that spatial modeling significantly improves the joint calibration of the standard (non-spatial) postprocessing methods. However, the slight bump at the left side of the histograms for spatial NGR$_{\textup c}$, ECC NGR$_+$ and ECC NGR$_{\textup c}$ suggest that for this specific region, the correlation under these models is slightly too high over longer distances. The opposite holds for spatial BMA and spatial NGR$_+$, where the observed field disproportionally often takes either the lowest or the highest rank. Overall, the multivariate methods still yield histograms which are much closer to uniformity than the univariate methods.

Table~\ref{tab:ScoresSaar} shows the multivariate scores over this region, and while these results are subject to some sampling variability, they show again a clear tendency of the spatial models yielding better multivariate performance in respect to their univariate counterparts with spatial NGR$_{\textup c}$ being especially competitive. The somewhat counter-intuitive result that the energy scores of ECC NGR$_{\textup c}$ and ECC NGR$_+$ are larger than those of NGR$_{\textup c}$ and NGR$_+$ might be explained by sampling effects. While the two latter are based on 10,000 samples, the ECC ensembles and the raw ensemble consist of only 20 members.  This does not warrant a stable estimation of the empirical covariance matrix which can be disastrous when calculating the Dawid-Sebastiani score and also have a negative impact on the energy score.  On the other hand, it illustrates that it can be problematic in certain contexts that ECC NGR$_+$ and ECC NGR$_{\textup c}$ inherit the sometimes close to singular correlation matrices from the raw COSMO-DE ensemble forecasts.

\begin{table}
  \centering
  \caption{Average energy score (ES), Euclidean error (EE) and Dawid-Sebastiani score (DS) of joint temperature forecasts at seven observation stations in the state of Saarland in Germany over all 346 days in the test set.}
  \label{tab:ScoresSaar}
  \smallskip
  \begin{tabular}{ld{1.2}d{1.2}d{5.1}}
    \toprule 
    & \multicolumn{1}{c}{\textup{ES}} & \multicolumn{1}{c}{\textup{EE}} & \multicolumn{1}{c}{\textup{DS}} \\
    & \multicolumn{1}{c}{\textup{(\textcelsius)}} & \multicolumn{1}{c}{\textup{(\textcelsius)}}\\
    \midrule 
    Raw ensemble              & 5.27 & 5.86 & 21210.9 \\ 
    BMA                       & 3.59 & 4.90 &    20.6 \\ 
    NGR$_+$                   & 3.59 & 4.91 &    20.7 \\ 
    NGR$_{\textup c}$         & 3.28 & 4.55 &    17.4 \\ 
    Spatial BMA               & 3.56 & 4.90 &    16.8 \\ 
    Spatial NGR$_+$           & 3.57 & 4.91 &    16.9 \\ 
    Spatial NGR$_{\textup c}$ & 3.25 & 4.55 &    14.2 \\ 
    ECC NGR$_+$               & 3.69 & 4.92 &  1112.9 \\ 
    ECC NGR$_{\textup c}$     & 3.36 & 4.55 &  1665.1 \\ 
    \bottomrule  
  \end{tabular}
\end{table}

\subsection{Case study II: Minimum temperature along the highway A3}

\begin{figure}
\begin{centering}
\includegraphics[scale=0.9]{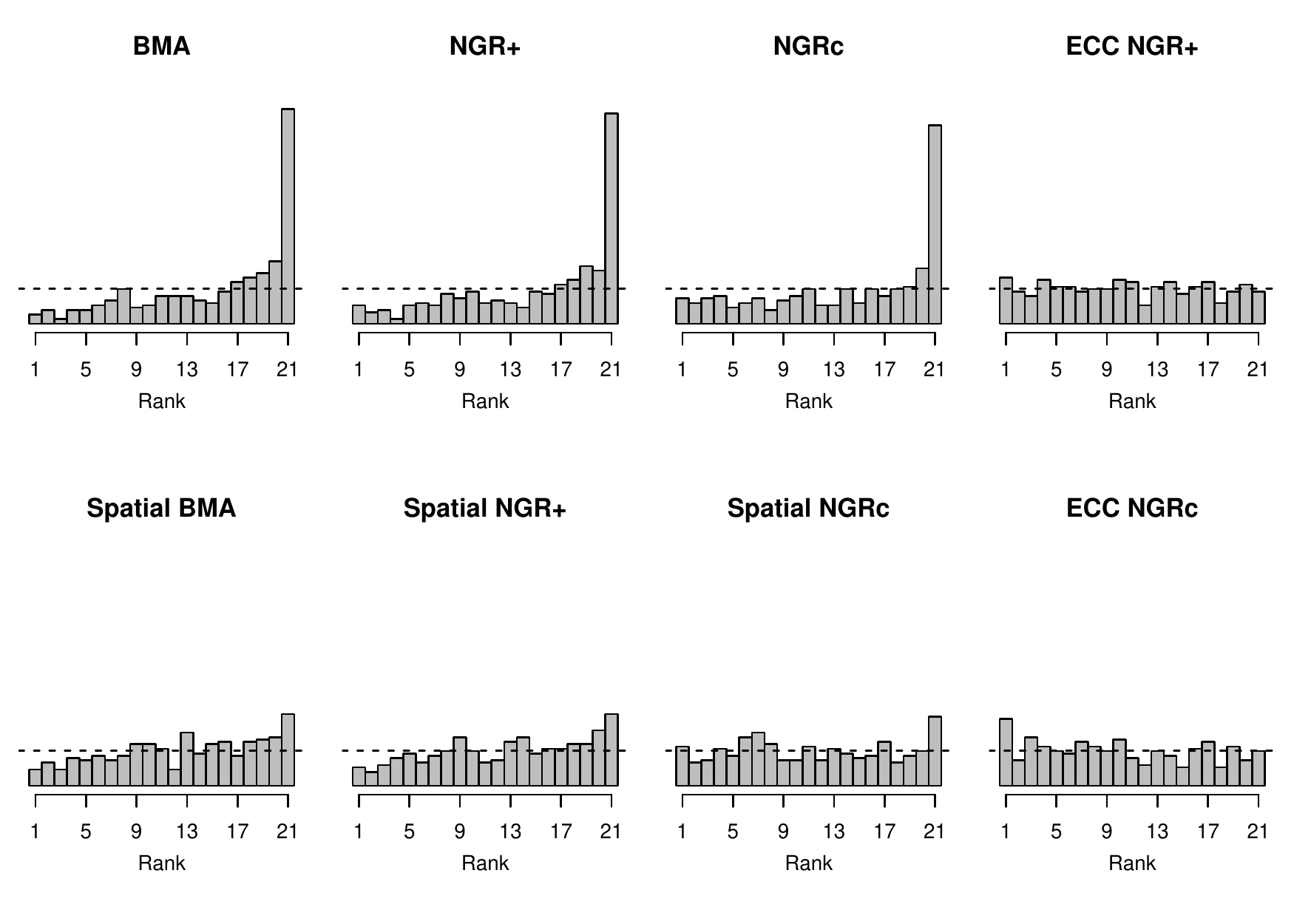}
\par\end{centering}
\caption{Verification rank histograms for forecasts of the minimum temperature over 11 stations along a section of the highway A3, where the different cases correspond to the days of the verification period.}
\label{fig:a3}
\end{figure}

As a second example in which the multivariate aspect of the predictive distributions becomes noticeable, we consider the task of predicting the minimum temperature along a section of the highway A3 which connects the two cities Frankfurt am Main and Cologne. For consistency with the forecasts at the individual stations and with other composite quantities, we do not set up a separate postprocessing model for minimum temperature, but derive it by taking the minimum over 11 stations along this section of the A3.  

Since the minimum of several random variables depends not only on their means and variances, but also on their correlations, we expect that only the spatial postprocessing methods can provide reasonably calibrated probabilistic forecasts.  Indeed, the histograms in Figure~\ref{fig:a3} show that without spatial modeling the minimum temperature is systematically underestimated. This is a consequence of the fact that the minimum over independent random variables is on average much smaller than the minimum over positively correlated random variables.  This systematic underestimation is largely avoided by spatial BMA, spatial NGR$_+$ and spatial NGR$_{\textup c}$ while the ECC techniques here yield the histograms closest to uniformity.  This clear advantage of postprocessing methods that account for spatial correlations is further confirmed by the CRPS and MAE scores in Table~\ref{tab:A3MinTemp}.

As an application and example of the relevance of spatial modeling in practice, consider the decision problem of dispatching or not dispatching salt spreaders when the temperatures along the considered section of the A3 are predicted to fall below 0\textcelsius. The event ``temperature falls below 0\textcelsius ~ at at least one location along the A3'' is equivalent to ``minimum temperature along the A3 falls below 0\textcelsius'', and good decisions are therefore taken if this event is predicted accurately. The last column of Table~\ref{tab:A3MinTemp} shows the corresponding average Brier scores (BS) over the verification days in the winter months of January, February and November, and illustrates once again that appropriate consideration of spatial dependence is required to take full advantage of statistical postprocessing.

\begin{table}
 \centering
  \caption{CRPS and MAE for minimum temperature forecasts over 11 stations along the highway A3 averaged over all verification days. The last column gives the BS for the event that the temperature drops below freezing (0\textcelsius) at at least one of these stations. The average is calculated over the winter subset of verification days in January, February and November 2011.}
  \label{tab:A3MinTemp}
  \smallskip
  \begin{tabular}{lccc}
    \toprule 
    & CRPS & MAE & BS$_0$\\
    & (\textcelsius) & (\textcelsius) & \\
    \midrule 
    Raw ensemble              & 1.74 & 1.92 & 0.120 \\
    BMA                       & 1.08 & 1.41 & 0.121 \\               
    NGR$_+$                   & 1.05 & 1.37 & 0.114 \\
    NGR$_{\textup c}$         & 0.96 & 1.27 & 0.103 \\
    Spatial BMA               & 0.86 & 1.20 & 0.079 \\
    Spatial NGR$_+$           & 0.86 & 1.20 & 0.082 \\
    Spatial NGR$_{\textup c}$ & 0.82 & 1.16 & 0.079 \\
    ECC NGR$_+$               & 0.84 & 1.18 & 0.085 \\
    ECC NGR$_{\textup c}$     & 0.84 & 1.17 & 0.083 \\
    \bottomrule  
  \end{tabular}
 \end{table}

\section{Discussion}\label{sec:6}

In this paper we have proposed a postprocessing method for temperature that uses the information of a dynamical ensemble as inputs and generates a calibrated statistical ensemble as an output. In doing so, it not only yields calibrated marginal predictive distributions but entire temperature forecast fields, thus aiming for multivariate calibration. The importance of this property is underlined by the results presented in Section \ref{sec:5} where forecasts of spatially aggregated quantities are studied and spatial correlations have to be considered. Our spatial NGR$_+$ approach performs similar to the spatial BMA approach of \citet{Berrocal2007}. However, it is conceptually simpler and computationally more efficient; the estimation of the spatial correlation structure of spatial BMA is $M$ times more expensive than that of spatial NGR$_+$, where $M$ is the size of the original ensemble. This makes it an attractive alternative, especially since further extensions -- such as the spatial NGR$_{\textup c}$ 
method presented here -- are also easier to implement.

In our case study using the ensemble forecasts of the COSMO-DE-EPS, the performance of the parametric spatial methods was overall slightly better than the results obtained by modeling spatial dependence via ECC. This result does not hold in any case. When the (spatial) correlation structure of the ensemble represents the true multivariate uncertainty well, methods that use or retain the rank correlations \citep{RoulinVannitsem2011,Schefzik&2013,VanSchaeybroeckVannitsem2013} have the potential advantage that they can feature flow-dependent dependence structures while the statistical models presented here rely on the assumption that correlations are constant over a certain period of time. A statistical approach, on the other hand, has the advantage that it determines the correlation structure based on both forecasts and observations, and thus does not inherit (or even amplify) spurious and wrong correlations that may be present in the ensemble.

The exponential correlation function used by \citet{Gel&2004}, \citet{Berrocal2007}, and in the present paper is of course a somewhat simplistic model. While replacing it by a function from the more general Mat\'ern class, which nests the exponential model as a special case, did not improve the performance of our method, Figure \ref{fig:differencePITs} suggests that a non-stationary correlation function might yield a better approximation of the true spatial dependence structure. There are a number of non-parametric modeling approaches that can potentially deal with these kinds of effects \citep{AnderesStein2011,Lindgren&2011,Jun&2011,Kleiber&2013}. However, this is rather challenging and left for future research.  A further extension of the approach presented here concerns correlations between different lead times. Instead of modeling spatial correlations only one would need to set up a model that captures correlations in both space and time.  Similarly, some applications require appropriate correlations 
between different weather variables. This presents yet another multivariate aspect that has been addressed by \citet{Moeller&2013}.  Taking all three aspects -- space, time, and different variables -- into account would be the ultimate goal in multivariate modeling. At the same time, this further increases the level of complexity so that in this very general setting the ECC approach might be preferred just for the sake of simplicity.

\section*{Acknowledgments}

The authors thank Tilmann Gneiting for sharing his thoughts and expertise. This work was funded by the German Federal Ministry of Education and Research, within the framework of the extramural research program of Deutscher Wetterdienst and by Statistics for Innovation, sfi$^2$ in Oslo, Norway. 

\bibliography{sEMOS}

\begin{thebibliography}{}

\bibitem[\protect\citeauthoryear{Anderes}{Anderes}{2011}]{AnderesStein2011}
Anderes, E.~and~Stein, M.~L. (2011).
\newblock Local likelihood estimation for nonstationary random fields.
\newblock {\em J Multivariate Anal\/}~{\em 102}, 506--520.

\bibitem[\protect\citeauthoryear{Anderson}{Anderson}{1996}]{Anderson1996}
Anderson, J.~L. (1996).
\newblock A method for producing and evaluating probabilistic forecasts from
  ensemble model integrations.
\newblock {\em J Climate\/}~{\em 9\/}(7), 1518--1530.

\bibitem[\protect\citeauthoryear{Baldauf, Seifert, F\"{o}rstner, Majewski,
  Raschendorfer, and Reinhardt}{Baldauf et~al.}{2011}]{Baldauf2011}
Baldauf, M., A.~Seifert, J.~F\"{o}rstner, D.~Majewski, M.~Raschendorfer, and
  T.~Reinhardt (2011).
\newblock {Operational convective-scale numerical weather prediction with the
  COSMO model: description and sensitivities}.
\newblock {\em Mon Weather Rev\/}~{\em 139\/}(12), 3887--3905.

\bibitem[\protect\citeauthoryear{Berrocal, Raftery, and Gneiting}{Berrocal
  et~al.}{2007}]{Berrocal2007}
Berrocal, V.~J., A.~E. Raftery, and T.~Gneiting (2007).
\newblock Combining spatial statistical and ensemble information in
  probabilistic weather forecasts.
\newblock {\em Mon Weather Rev\/}~{\em 135\/}(4), 1386--1402.

\bibitem[\protect\citeauthoryear{Brier}{Brier}{1950}]{Brier1950}
Brier, G.~W. (1950).
\newblock Verification of forecasts expressed in terms of probability.
\newblock {\em Mon Weather Rev\/}~{\em 78}, 1--3.

\bibitem[\protect\citeauthoryear{Br\"ocker}{Br\"ocker}{2012}]{Broecker2012}
Br\"ocker, J. (2012).
\newblock Evaluating raw ensembles with the continous ranked probability score.
\newblock {\em Q J Roy Meteor Soc\/}~{\em 138}, 1611--1617.

\bibitem[\protect\citeauthoryear{Br\"{o}cker and Smith}{Br\"{o}cker and
  Smith}{2008}]{BroeckerSmith2008}
Br\"{o}cker, J. and L.~A. Smith (2008).
\newblock {From ensemble forecasts to predictive distribution functions}.
\newblock {\em Tellus A\/}~{\em 60}, 663--678.

\bibitem[\protect\citeauthoryear{Bryd, Lu, Nocedal, and Zhu}{Bryd
  et~al.}{1995}]{Bryd1995}
Bryd, R.~H., P.~Lu, J.~Nocedal, and C.~Zhu (1995).
\newblock A limited memory algorithm for bound constrained optimization.
\newblock {\em SIAM J Sci Comp\/}~{\em 16}, 1190--1208.

\bibitem[\protect\citeauthoryear{Cressie}{Cressie}{1985}]{Cressie1985}
Cressie, N. A.~C. (1985).
\newblock Fitting variogram models by weighted least squares.
\newblock {\em Math Geol\/}~{\em 17}, 563--586.

\bibitem[\protect\citeauthoryear{Dawid}{Dawid}{1984}]{Dawid1984}
Dawid, A.~P. (1984).
\newblock Statistical theory: The prequential approach (with discussion and
  rejoinder).
\newblock {\em J Roy Stat Soc A\/}~{\em 147}, 278--292.

\bibitem[\protect\citeauthoryear{Dawid and Sebastiani}{Dawid and
  Sebastiani}{1999}]{DawidSebastiani1999}
Dawid, A.~P. and P.~Sebastiani (1999).
\newblock Coherent dispersion criteria for optimal experimental design.
\newblock {\em Ann Stat\/}~{\em 27}, 65--81.

\bibitem[\protect\citeauthoryear{Delle~Monache, Hacker, Y., X., and
  B.}{Delle~Monache et~al.}{2006}]{DelleMonache&2006}
Delle~Monache, L., J.~P. Hacker, Z.~Y., D.~X., and S.~R. B. (2006).
\newblock Probabilistic aspects of meteorological and ozone regional ensemble
  forecasts.
\newblock {\em J Geophys Res\/}~{\em 111}, D24307.

\bibitem[\protect\citeauthoryear{Doms and Sch{\"a}ttler}{Doms and
  Sch{\"a}ttler}{2002}]{DomsSchaettler2002}
Doms, G. and U.~Sch{\"a}ttler (2002).
\newblock A description of the nonhydrostatic regional model {LM} : Dynamics
  and numerics.
\newblock Technical report, Deutscher Wetterdienst.

\bibitem[\protect\citeauthoryear{Gebhardt, Theis, Paulat, and
  Ben-Bouall\`{e}gue}{Gebhardt et~al.}{2011}]{Gebhardt2011}
Gebhardt, C., S.~E. Theis, M.~Paulat, and Z.~Ben-Bouall\`{e}gue (2011).
\newblock Uncertainties in {COSMO-DE} precipitation forecasts introduced by
  model perturbations and variation of lateral boundaries.
\newblock {\em Atmos Res\/}~{\em 100}, 168--177.

\bibitem[\protect\citeauthoryear{Gel, Raftery, and Gneiting}{Gel
  et~al.}{2004}]{Gel&2004}
Gel, Y., A.~E. Raftery, and T.~Gneiting (2004).
\newblock {Calibrated probabilistic mesoscale weather field forecasting: The
  geostatistical output perturbation (GOP) method (with discussion and
  rejoinder)}.
\newblock {\em J Am Stat Assoc\/}~{\em 99}, 575--590.

\bibitem[\protect\citeauthoryear{Gneiting}{Gneiting}{2011}]{Gneiting2011}
Gneiting, T. (2011).
\newblock Making and evaluating point forecasts.
\newblock {\em J Am Stat Assoc\/}~{\em 106\/}(494), 746--762.

\bibitem[\protect\citeauthoryear{Gneiting, Balabdaoui, and Raftery}{Gneiting
  et~al.}{2007}]{Gneiting2007a}
Gneiting, T., F.~Balabdaoui, and A.~E. Raftery (2007).
\newblock {Probabilistic forecasts, calibration and sharpness}.
\newblock {\em J Roy Stat Soc B\/}~{\em 69\/}(2), 243--268.

\bibitem[\protect\citeauthoryear{Gneiting and Raftery}{Gneiting and
  Raftery}{2007}]{Gneiting2007}
Gneiting, T. and A.~E. Raftery (2007).
\newblock {Strictly proper scoring rules, prediction, and estimation}.
\newblock {\em J Am Stat Assoc\/}~{\em 102\/}(477), 359--378.

\bibitem[\protect\citeauthoryear{Gneiting, Raftery, Westveld, and
  Goldman}{Gneiting et~al.}{2005}]{Gneiting2005}
Gneiting, T., A.~E. Raftery, A.~H. Westveld, and T.~Goldman (2005).
\newblock {Calibrated probabilistic forecasting using ensemble model output
  statistics and minimum CRPS estimation}.
\newblock {\em Mon Weather Rev\/}~{\em 133\/}(5), 1098--1118.

\bibitem[\protect\citeauthoryear{Grimit, Gneiting, Berrocal, and
  Johnson}{Grimit et~al.}{2006}]{Grimit&2006}
Grimit, E.~P., T.~Gneiting, V.~J. Berrocal, and N.~A. Johnson (2006).
\newblock {The continuous ranked probability score for circular variables and
  its application to mesoscale forecast ensemble verification}.
\newblock {\em Q J Roy Meteor Soc\/}~{\em 132}, 2925--2942.

\bibitem[\protect\citeauthoryear{Hagedorn, Hamill, and Whitaker}{Hagedorn
  et~al.}{2008}]{Hagedorn&2008}
Hagedorn, R., T.~M. Hamill, and J.~S. Whitaker (2008).
\newblock Probabilistic forecast calibration using {ECMWF} and {GFS} ensemble
  reforecasts. {P}art {I}: Two-meter temperatures.
\newblock {\em Mon Weather Rev\/}~{\em 136}, 2608--2619.

\bibitem[\protect\citeauthoryear{Hamill and Colucci}{Hamill and
  Colucci}{1997}]{Hamill1997}
Hamill, T.~M. and S.~J. Colucci (1997).
\newblock {Verification of Eta-RSM Short-Range Ensemble Forecasts}.
\newblock {\em Mon Weather Rev\/}~{\em 125\/}(6), 1312--1327.

\bibitem[\protect\citeauthoryear{Jun, Szunyogh, Genton, Zhang, and Bishop}{Jun
  et~al.}{2011}]{Jun&2011}
Jun, M., I.~Szunyogh, M.~G. Genton, F.~Zhang, and C.~H. Bishop (2011).
\newblock A statistical investigation of the sensitivity of ensemble-based
  {K}alman filters to covariance filtering.
\newblock {\em Mon Weather Rev\/}~{\em 139\/}(9), 3036--3051.

\bibitem[\protect\citeauthoryear{Kann, Wittmann, Wang, and Ma}{Kann
  et~al.}{2009}]{Kann&2009}
Kann, A., C.~Wittmann, Y.~Wang, and X.~Ma (2009).
\newblock Calibrating 2-m temperature of limited-area ensemble forecasts using
  high-resolution analysis.
\newblock {\em Mon Weather Rev\/}~{\em 137}, 3373--3387.

\bibitem[\protect\citeauthoryear{Kleiber, Katz, and Rajagopalan}{Kleiber
  et~al.}{2013}]{Kleiber&2013}
Kleiber, W., R.~Katz, and B.~Rajagopalan (2013).
\newblock Daily minimum and maximum temperature simulation over complex
  terrain.
\newblock {\em Ann Appl Stat\/}~{\em 7}, 588--612.

\bibitem[\protect\citeauthoryear{Kleiber, Raftery, Baars, Gneiting, Mass, and
  Grimit}{Kleiber et~al.}{2011}]{Kleiber2011}
Kleiber, W., A.~E. Raftery, J.~Baars, T.~Gneiting, C.~F. Mass, and E.~P. Grimit
  (2011).
\newblock {Locally calibrated probabilistic temperature forecasting using
  geostatistical model averaging and local Bayesian model averaging}.
\newblock {\em Mon Weather Rev\/}~{\em 139\/}(570), 2630--2649.

\bibitem[\protect\citeauthoryear{Lerch and Thorarinsdottir}{Lerch and
  Thorarinsdottir}{2013}]{LerchThorarinsdottir2013}
Lerch, S. and T.~L. Thorarinsdottir (2013).
\newblock Comparison of nonhomogeneous regression models for probabilistic wind
  speed forecasting.
\newblock {\em Tellus A\/}~{\em 65}, 21206.

\bibitem[\protect\citeauthoryear{Leutbecher and Palmer}{Leutbecher and
  Palmer}{2008}]{LeutbecherPalmer2008}
Leutbecher, M. and T.~N. Palmer (2008).
\newblock Ensemble forecasting.
\newblock {\em J Comput Phys\/}~{\em 227}, 3515--3539.

\bibitem[\protect\citeauthoryear{Lewis}{Lewis}{2005}]{Lewis2005}
Lewis, J.~M. (2005).
\newblock {Roots of Ensemble Forecasting}.
\newblock {\em Mon Weather Rev\/}~{\em 133}, 1865--1885.

\bibitem[\protect\citeauthoryear{Lindgren, Rue, and Lindstr\"om}{Lindgren
  et~al.}{2011}]{Lindgren&2011}
Lindgren, F., H.~Rue, and J.~Lindstr\"om (2011).
\newblock An explicit link between {G}aussian fields and {G}aussian {M}arkov
  random fields: The stochastic partial differential equation approach'' (with
  discussion).
\newblock {\em J Roy Stat Soc B\/}~{\em 73}, 423--498.

\bibitem[\protect\citeauthoryear{M{\"o}ller, Lenkoski, and
  Thorarinsdottir}{M{\"o}ller et~al.}{2013}]{Moeller&2013}
M{\"o}ller, A., A.~Lenkoski, and T.~L. Thorarinsdottir (2013).
\newblock Multivariate probabilistic forecasting using {B}ayesian model
  averaging and copulas.
\newblock {\em Q J Roy Meteor Soc\/}~{\em 139\/}(673), 982--991.

\bibitem[\protect\citeauthoryear{Peralta and Buchhold}{Peralta and
  Buchhold}{2011}]{Peralta2011}
Peralta, C. and M.~Buchhold (2011).
\newblock Initial condition perturbations for the {COSMO-DE-EPS}.
\newblock {\em COSMO Newsletter\/}~{\em 11}, 115--123.

\bibitem[\protect\citeauthoryear{Pinson and Tastu}{Pinson and
  Tastu}{2013}]{PinsonTastu2013}
Pinson, P. and J.~Tastu (2013).
\newblock Discrimination ability of the energy score.
\newblock Technical report, Technical University of Denmark.

\bibitem[\protect\citeauthoryear{{R Core Team}}{{R Core Team}}{2013}]{R2013}
{R Core Team} (2013).
\newblock {\em R: A Language and Environment for Statistical Computing}.
\newblock Vienna, Austria: R Foundation for Statistical Computing.

\bibitem[\protect\citeauthoryear{Raftery, Gneiting, Balabdaoui, and
  Polakowski}{Raftery et~al.}{2005}]{Raftery2005}
Raftery, A.~E., T.~Gneiting, F.~Balabdaoui, and M.~Polakowski (2005).
\newblock {Using Bayesian model averaging to calibrate forecast ensembles}.
\newblock {\em Mon Weather Rev\/}~{\em 133\/}(5), 1155--1174.

\bibitem[\protect\citeauthoryear{Rasmussen and Williams}{Rasmussen and
  Williams}{2006}]{RasmussenWilliams2006}
Rasmussen, C.~E. and C.~K.~I. Williams (2006).
\newblock {\em Gaussian processes for machine learning}.
\newblock Cambridge, MA: The MIT Press.

\bibitem[\protect\citeauthoryear{Roulin and Vannitsem}{Roulin and
  Vannitsem}{2011}]{RoulinVannitsem2011}
Roulin, E. and S.~Vannitsem (2011).
\newblock Post-processing of ensemble precipitation predictions with extended
  logistic regression based on hindcasts.
\newblock {\em Mon Weather Rev\/}~{\em 47}, 874--888.

\bibitem[\protect\citeauthoryear{Schefzik, Thorarinsdottir, and
  Gneiting}{Schefzik et~al.}{2013}]{Schefzik&2013}
Schefzik, R., T.~L. Thorarinsdottir, and T.~Gneiting (2013).
\newblock Uncertainty quantification in complex simulation models using
  ensemble copula coupling.
\newblock {\em Stat Sci\/}~{\em 28\/}(4), 616--640.

\bibitem[\protect\citeauthoryear{Scheuerer}{Scheuerer}{2014}]{Scheuerer2013}
Scheuerer, M. (2014).
\newblock Probabilistic quantitative precipitation forecasting using ensemble
  model output statistics.
\newblock {\em Q J Roy Meteor Soc\/}~{\em 140}, 1086--1096.

\bibitem[\protect\citeauthoryear{Scheuerer and B\"uermann}{Scheuerer and
  B\"uermann}{2014}]{ScheuererBuermann2013}
Scheuerer, M. and L.~B\"uermann (2014).
\newblock Spatially adaptive post-processing of ensemble forecasts for
  temperature.
\newblock {\em J Roy Stat Soc C\/}~{\em 63}, 405--422.

\bibitem[\protect\citeauthoryear{Scheuerer and K\"onig}{Scheuerer and
  K\"onig}{2014}]{ScheuererKoenig2013}
Scheuerer, M. and G.~K\"onig (2014).
\newblock Gridded locally calibrated, probabilistic temperature forecasts based
  on ensemble model output statistics.
\newblock {\em Q J Roy Meteor Soc\/}.
\newblock To appear. DOI:10.1002/qj.2323.

\bibitem[\protect\citeauthoryear{Schlather}{Schlather}{2011}]{Schlather2011}
Schlather, M. (2011).
\newblock {RandomFields: Simulation and Analysis of Random Fields}.

\bibitem[\protect\citeauthoryear{Steppeler, Doms, Sch\"{a}ttler, Bitzer,
  Gassmann, Damrath, and Gregoric}{Steppeler et~al.}{2003}]{Steppeler2003}
Steppeler, J., G.~Doms, U.~Sch\"{a}ttler, H.~W. Bitzer, A.~Gassmann,
  U.~Damrath, and G.~Gregoric (2003).
\newblock Meso-gamma scale forecasts using the nonhydrostatic model {LM}.
\newblock {\em Meteorol Atmos Phys\/}~{\em 82}, 75--96.

\bibitem[\protect\citeauthoryear{Thorarinsdottir and Gneiting}{Thorarinsdottir
  and Gneiting}{2010}]{Thorarinsdottir2010}
Thorarinsdottir, T.~L. and T.~Gneiting (2010).
\newblock Probabilistic forecasts of wind speed: Ensemble model output
  statistics by using heteroscedastic censored regression.
\newblock {\em J Roy Stat Soc A\/}~{\em 173\/}(2), 371--388.

\bibitem[\protect\citeauthoryear{Thorarinsdottir and Johnson}{Thorarinsdottir
  and Johnson}{2012}]{Thorarinsdottir2012}
Thorarinsdottir, T.~L. and M.~S. Johnson (2012).
\newblock Probabilistic wind gust forecasting using non-homogeneous gaussian
  regression.
\newblock {\em Mon Weather Rev\/}~{\em 140}, 889--897.

\bibitem[\protect\citeauthoryear{Thorarinsdottir, Scheuerer, and
  Heinz}{Thorarinsdottir et~al.}{2013}]{Thorarinsdottir&2013b}
Thorarinsdottir, T.~L., M.~Scheuerer, and C.~Heinz (2013).
\newblock Assessing the calibration of high-dimensional ensemble forecasts
  using rank histograms.
\newblock arXiv:1310.0236.

\bibitem[\protect\citeauthoryear{Van~Schaeybroeck and
  Vannitsem}{Van~Schaeybroeck and
  Vannitsem}{2013}]{VanSchaeybroeckVannitsem2013}
Van~Schaeybroeck, B. and S.~Vannitsem (2013).
\newblock Ensemble post-processing using member-by-member approaches.
\newblock In submission.

\bibitem[\protect\citeauthoryear{Wilks}{Wilks}{2011}]{Wilks2011}
Wilks, D.~S. (2011).
\newblock {\em {Statistical Methods in the Atmospheric Sciences}\/} (3rd ed.).
\newblock Elsevier Academic Press, Amsterdam.

\bibitem[\protect\citeauthoryear{Wilks and Hamill}{Wilks and
  Hamill}{2007}]{WilksHamill2007}
Wilks, D.~S. and T.~M. Hamill (2007).
\newblock Comparison of ensemble-{MOS} methods using {GFS} reforecasts.
\newblock {\em Mon Weather Rev\/}~{\em 135}, 2379--2390.

\end{thebibliography}

\appendix

\section{Forecast evaluation methods}

Statistical postprocessing aims at correcting systematic biases and/or misrepresentation of the forecast uncertainty in the raw ensemble and, in our case, returns full probabilistic distributions.  To evaluate the predictive performance of the methods under consideration, we follow \cite{Gneiting2007a} who state that the goal of probabilistic forecasting is to maximize the sharpness of the predictive distribution subject to calibration.  

\subsection{Assessing calibration}

Calibration refers to the statistical compatibility between the forecasts and the observations; the forecast is calibrated if the observation cannot be distinguished from a random draw from the predictive distribution. For continuous univariate distributions, calibration can be assessed empirically by plotting the histogram of the probability integral transform (PIT) -- the value of the predictive cumulative distribution function in the observed value \citep{Dawid1984, Gneiting2007a} -- over all forecast cases. A forecasting method that is calibrated on average will return a uniform histogram, a $\cap$-shape indicates overdispersion and a $\cup$-shape indicates underdispersion while a systematic bias results in a triangular shape histogram. The discrete equivalent of the PIT histogram, which applies to ensemble forecasts, is the verification rank histogram \citep{Anderson1996, Hamill1997}. It shows the distribution of the ranks of the observations within the corresponding ensembles and has the same interpretation 
as the PIT histogram. In order to facilitate direct comparison of the various methods, we only employ the rank histogram.  That is, for the continuous predictive distributions, we create a $20$-member ensemble given by $20$ random samples from the distribution. 

For multivariate settings, we employ the band depth rank histogram proposed by \cite{Thorarinsdottir&2013b}.  This approach ranks the observation within a sample of forecast scenarios by assessing the centrality of the observation within the sample. Let $\mathbf{X} = \{ \mathbf{x}_1, \ldots, \mathbf{x}_{M+1} \} = \{ \mathbf{Y}, \mathbf{F}_1, \ldots, \mathbf{F}_M\}$ denote a set of $M$ forecast vectors and the observation $\mathbf{Y}$, each of dimension $d$. To calculate the band depth rank of the observation $\mathbf{Y}$ in $\mathbf{X}$, we first apply the pre-rank function 
\begin{align*}
r (\mathbf{x}) & = \frac{1}{d} \sum_{k=1}^d \sum_{1 \leq i_1 < i_2 \leq M + 1} \mathbf{1}  \big\{ \min \{x_{i_1 k}, x_{i_2 k} \} \leq x_k \leq \max \{ x_{i_1 k}, x_{i_2 k}\}\big\} \\
& = \frac{1}{d} \sum_{k=1}^d \big[ M + 1 - \textup{rank}_{\mathbf{X}}(x_k) \big] \big[ \textup{rank}_{\mathbf{X}}(x_k) - 1 \big] + M,
\end{align*}
to all vectors $\mathbf{x} \in \mathbf{X}$, where $\mathbf{1}\{ \cdot \}$ denotes the indicator function and $\textup{rank}_{\mathbf{X}}(x_k)$ denote the rank of element $k$ of the vector $\mathbf{x}$ in $\mathbf{X}$. The band depth rank of $\mathbf{x}_i$ is then given by the rank of $r(\mathbf{x}_i)$ in $\{ r(\mathbf{x}_1), \ldots, r(\mathbf{x}_{M+1}) \}$ with ties resolved at random. Calibrated forecasts should result in a uniform histogram. However, the interpretation of miscalibration in the band depth rank histogram is somewhat different than that of the classic univariate rank histogram.  A skew histogram with too many high ranks is an indication of an overdispersive ensemble while to many low ranks can result from either an underdispersive or biased ensemble. Furthermore, too high correlations in the ensemble produce a $\cap$-shaped histogram while a $\cup$-shaped histogram is an indication of a lack of correlation in the ensemble. 

Alternatively, we also investigate the fit of the correlation structure by investigating the calibration of predicted temperature differences at close-by stations. Under the multivariate predictive distribution model in (\ref{eq:spatialNGR}), the predictive distribution of the temperature difference between location $s_i$ and $s_j$ is given by
\bl
\begin{equation}\label{eq:PrdDstb-tempdiff}
(\Delta y)_{s_is_j}|f_{1s_i},f_{1s_j},...,f_{Ms_i},f_{Ms_j},\mathcal{D}_{\mathcal{S}}^{\T} \sim \mathcal{N}\big(\mu_{s_i}-\mu_{s_j}, \sigma_{s_i}^2-2\rho_{s_is_j}\sigma_{s_i}\sigma_{s_j}+\sigma_{s_j}^2\big),
\end{equation}
\el
where $\mu_{s_i},\mu_{s_j}$ and $\sigma_{s_i}^2,\sigma_{s_j}^2$ are the predictive means and variances at $s_i$ and $s_j$ and $\rho_{s_is_j}=C_{\theta,r}\left(s_i,s_j\right)$ is the correlation between the forecast errors at those two locations.  For each station, we calculate the observed temperature differences between this station and all stations within a radius of $50$ km, and calculate the PIT values of the predictive distributions given by (\ref{eq:PrdDstb-tempdiff}).  In the absence of a spatial model, we take $\rho_{s_is_j}=0$ for all $s_i,s_j\in\mathcal{S}$.  For a combination of ECC and NGR where the multivariate distribution is represented by an ensemble, we approximate (\ref{eq:PrdDstb-tempdiff}) by the empirical CDF of the temperature differences predicted by the individual ensemble members.  

Assuming no local biases and marginal calibration, the calibration of the temperature difference forecasts mainly depends on the correct specification of $\rho_{s_is_j}$.  It will be underdispersive if $\rho_{s_is_j}$ is overestimated and overdispersive if $\rho_{s_is_j}$ is underestimated. If the strength of spatial correlations implied by the respective postprocessing approach is adequate, the predictive distributions of temperature differences are calibrated and the corresponding PIT values are uniformly distributed on $[0,1]$.  Underestimating the correlation strength would entail $\cap$-shaped PIT histograms, i.e.\ PIT values would tend to accumulate around $0.5$.  Conversely, overestimating the correlation strength would yield PIT values closer to $0$ or $1$.  A station-specific PIT histogram may thus be summarized by the mean absolute deviations of the PIT values from $0.5$ over all verification days and all temperature differences between this station and stations within the $50$ km neighborhood.

The information provided by a rank histogram may also be summarized numerically by the reliability index (RI) which is defined as
\[
 \mathrm{RI} = \sum_{i = 1}^I \Big| \zeta_i - \frac{1}{I} \Big|,
\]
where $I$ is the number of (equally-sized) bins in the histogram and $\zeta_i$ is the observed relative frequency in bin $i = 1, \ldots, I$.  The reliability index thus measures the departure of the rank histogram from uniformity \citep{DelleMonache&2006}. 

\subsection{Scoring rules}

While rank histograms are a useful calibration diagnostic tool, they do not yield information on the sharpness of the predictive distributions.  The latter can be evaluated by studying the average width of prediction intervals, which should be as small as possible, provided that the empirical coverage is close to the nominal coverage.  As a quantitative measure for predictive performance that takes both calibration and sharpness into account, we employ several proper scoring rules \citep{Gneiting2007}.  The different scores assess different aspects of the forecasts. However, they are all negatively oriented in that a smaller score indicates a better forecast.  For events with a binary outcome, e.g.\ ``the temperature $y$ does not exceed a certain threshold $x$'', we use the Brier score \citep{Brier1950}
\[
\textup{bs}_x(G, y) = \big({\bf 1}_{\{y\leq x\}}-G(x) \big)^2
\]
where ${\bf 1}_{\{y\leq x\}}$ is equal to one if $y\leq x$ and zero otherwise, and $G(x)$ is the predicted probability for $y\leq x$.  

The continuous ranked probability score (CRPS) in \eqref{eq:crps} is the integral of the Brier scores over all thresholds $x\in\real$ and thus an overall performance measure.  When the integral in \eqref{eq:crps} is not available in a closed form, the equivalent formulation 
\bl
\begin{equation}\label{eq:crps2}
\textup{crps}(G, y) = \mathbb{E} | X - y | - \frac{1}{2} \mathbb{E} | X - X' |
\end{equation}
\el
may be employed instead \citep{Gneiting2007}. Here, $\mathbb{E}$ denotes expectation, $| \cdot |$ stands for the absolute value and $X$ and $X'$ are independent copies of a random variable with cumulative distribution function $G$.  To estimate the expression in \eqref{eq:crps2}, we generate two independent samples ${\bf x} = \{ x_j\}_{j=1}^J$ and ${\bf x}' = \{x'_j \}_{j=1}^J$ from the predictive distribution and calculate
\bl
\[
\widehat{\textup{crps}}(G, y) = \sum_{j = 1}^J |x_j - y| - \frac{1}{2} \sum_{j=1}^J | x_j - x'_j|,  
\]
\el
where we typically set $J = 5000$.  The formulation in \eqref{eq:crps2} further permits the evaluation of the CRPS for discrete distributions such as an ensemble \citep{Grimit&2006}.  For multivariate distributions, the energy score (ES) provides a similar measure of predictive skill \citep{Gneiting2007}. It is given by 
\bl
\[ 
\textup{es}(G, \boldsymbol y) = \mathbb{E} \| \boldsymbol X - \boldsymbol y \| - \frac{1}{2} \mathbb{E} \| \boldsymbol X - \boldsymbol X' \|,
\]
\el
where $\| \cdot \|$ denotes the Euclidean norm, and may be approximated as the CRPS above. 

It has been noted \citep{PinsonTastu2013} that the sensitivity of the energy score to misrepresentation of the correlation structure is rather limited. As an additional score, we therefore consider the Dawid-Sebastiani (DS) score which depends on the predictive mean vector $\boldsymbol \mu_G$ and the predictive covariance matrix $\boldsymbol \Sigma_G$ of the multivariate predictive distribution $G$ via
\bl
\[ 
\textup{ds}(G, \boldsymbol y) = -\log\det \boldsymbol \Sigma_G - (\boldsymbol y- \boldsymbol \mu_G)^t\, \boldsymbol \Sigma_G^{-1}\,( \boldsymbol y - \boldsymbol \mu_G)
\]
\el
\citep{DawidSebastiani1999, Gneiting2007}. For multivariate Gaussian distributions such as the spatial NGR predictive distributions, the Dawid-Sebastiani score is equal to the logarithmic or ignorance score and may be calculated directly.  For spatial BMA, the mean and covariance matrix may be calculated as follows.  Let $\mathbf{Y}$ be a random vector which distribution is given by a mixture of $M$ Gaussian distributions each with mean $\boldsymbol \mu_m$, covariance $\boldsymbol \Sigma_m$ and weight $\omega_m$ for $m = 1, \ldots, M$. Then is holds that,
\[
\mathbb{E}(Y_i) = \sum_{m=1}^M \omega_m \mu_{mi} 
\]
and 
\[
\mathbb{E}(Y_i Y_j) = \sum_{m=1}^M \omega_m \big( (\boldsymbol \Sigma_m)_{ij} + \mu_{mi} \mu_{mj} \big).  
\]
The former formula can now  be used to calculate the mean $\boldsymbol \mu_G$ while the covariance matrix may be calculated by noting that $(\boldsymbol \Sigma_G)_{ij} = \mathbb E (Y_i Y_j) - \mathbb E (Y_i) \mathbb E (Y_j)$.  When $\Sigma_G$ must be estimated non-parametrically from a sample, such as for ECC, the calculations may be numerically unstable.  In this case, we add $0.00001$ to all elements on the diagonal in order to improve the numerical stability \citep{RasmussenWilliams2006}. 

Finally, we provide some error measures of the deterministic forecasts that are obtained as functionals (e.g.\ mean or median) of the predictive distributions.  For univariate probabilistic forecasts, the mean absolute error (MAE) and the root mean squared error (RMSE) assess the average proximity of the observation to the center of the predictive distribution.  The absolute error is calculated as the absolute difference between the observation and the median of the predictive distribution while the squared error is calculated using the mean of the predictive distribution \citep{Gneiting2011}.  The Euclidean error (EE) is the natural generalization of the absolute error to higher dimensions.  It is given by the Euclidean distance between the observation and the median of the predictive distribution.  The median of a multivariate predictive distribution is estimated using the functionality of the {\tt R}-package {\tt ICSNP}.

\end{document}